%% file: main_final.tex
\documentclass[lettersize,journal]{IEEEtran}

\usepackage{algorithm,algorithmic,amsmath,amssymb,amsthm,bbm,cite,color,comment,graphicx,microtype,setspace,url,caption}
\usepackage[USenglish]{babel}
\usepackage{hyperref}
\usepackage[utf8]{inputenc}
\usepackage[T1]{fontenc}
\usepackage[font=small]{caption}
\usepackage{lipsum}

\usepackage{enumitem}
\setitemize{itemsep=0mm,topsep=0mm,parsep=0pt,partopsep=0pt}
\usepackage{stfloats}
\usepackage{tikz}
\usetikzlibrary{arrows,backgrounds,calc,intersections,decorations.pathreplacing,positioning,shapes}
\makeatletter
\newcommand\subparagraph{%
  \@startsection{subparagraph}{5}
  {\parindent}
  {3.25ex \@plus 1ex \@minus .2ex}
  {-1em}
 {\normalfont\normalsize\bfseries}}
\makeatother
\usepackage{titlesec}
\let\subparagraph\relax

\usepackage{titlesec}

\usetikzlibrary{arrows, arrows.meta, backgrounds, calc, intersections, decorations.markings, decorations.pathreplacing, positioning, shapes}

\usepackage{pgfplots}
\usepgfplotslibrary{groupplots}
\pgfplotsset{compat=1.17}

\usepackage[font=small]{caption}

\input{./Definitions.tex}

\newcommand{\bs}{\textnormal{\tiny{AP}}}
\newcommand{\dl}{\textnormal{\tiny{DL}}}
\newcommand{\dlA}{\textnormal{\tiny{DL-1}}}
\newcommand{\dlB}{\textnormal{\tiny{DL-2}}}

\newcommand{\ul}{\textnormal{\tiny{UL}}}
\newcommand{\ulA}{\textnormal{\tiny{UL-1}}}
\newcommand{\ulB}{\textnormal{\tiny{UL-2}}}
\newcommand{\ulC}{\textnormal{\tiny{UL-3}}}
\newcommand{\ue}{\textnormal{\tiny{UE}}}
\newcommand{\mse}{\mathrm{MSE}}

\title{Over-the-Air Beamforming Desing for Full-Duplex Cell-Free Massive MIMO Systems}

\author{
Bikshapathi Gouda, and Antti T\"olli
\thanks{The authors are with the Centre for Wireless Communications, University of Oulu, Finland (e-mail: \{bikshapathi.gouda, antti.tolli\}@oulu.fi). 
Bikshapathi Gouda is currently with Skylo Technologies, Espoo, Finland. This work was conducted while he was with the University of Oulu.}
\vspace{-3mm}
}

\begin{document}
\maketitle

\begin{abstract}
We study a full-duplex (FD) cell-free massive MIMO system where distributed access points (APs) operate in FD mode while user equipments (UEs) remain half-duplex. Although simultaneous uplink (UL) and downlink (DL) transmissions improve spectral efficiency, they introduce residual self-interference, AP-to-AP coupling, and UE-to-UE cross-link interference. Building on prior over-the-air distributed beamforming frameworks, we develop a fully distributed beamforming design based on iterative UL and DL pilot signaling under a joint UL and DL sum mean-square error criterion that explicitly accounts for these interference components. In FD operation, simultaneous UL and DL pilot transmissions cause UE-to-UE pilot leakage, which contaminates the reconstruction of the cross terms required for AP-specific beamforming design. To mitigate this effect, we introduce a pilot-domain projection of the received signals at the UEs, which suppresses the interference component and enables accurate cross-term reconstruction at the APs. In addition, best-response updates at the UEs are employed within the alternating optimization framework to improve convergence under strong UE-to-UE interference. Numerical results demonstrate faster convergence and higher effective sum rate, with particularly significant gains for strongly interfering UEs, compared with both separate UL and DL distributed OTA beamforming training schemes and designs based solely on local channel state information.

\begin{IEEEkeywords}
Cell-free massive MIMO, distributed beamforming, full-duplex,   MSE, OTA signaling.
\end{IEEEkeywords}

\end{abstract}

\section{Introduction}

Cell-free massive multiple-input multiple-output (MIMO) has emerged as a key architectural paradigm for future wireless systems due to its ability to provide uniformly high spectral efficiency through coherent joint transmission (CJT) from geographically distributed access points (APs) \cite{Ngo2017CellFree,Nayebi2017PerformanceCF,Interdonato2019ScalableCF}. By eliminating conventional cell boundaries and enabling all APs to jointly serve user equipments (UEs), cell-free systems effectively mitigate the inter-cell interference and significantly enhance cell-edge UE performance. However, realizing the full benefits of CJT requires global channel state information (CSI) for beamforming design at the APs, which typically involves extensive fronthaul signaling and centralized processing \cite{Bjornson2019MakingCFScalable}. To address this scalability challenge, iterative bi-directional training (IBT) via over-the-air (OTA) signaling frameworks have been developed to enable fully distributed beamforming design without explicit CSI exchange, achieving performance comparable to centralized designs 
\cite{Rogalin2014ScalableBF,Atz21, Gouda2024MulticastOTA,Gouda2021DistributedRx,Shi2011WMMSE}.

Full-duplex (FD) operation at the APs is a promising extension of cell-free architectures, enabling simultaneous uplink (UL) and downlink (DL) transmission over the same time–frequency resource and potentially improving spectral efficiency \cite{Sabharwal2014FDReview,Bharadia2013FD,Everett2014PassiveSI}. However, FD operation introduces additional interference components, including residual self-interference (SI), AP-to-AP coupling, and UE-to-UE cross-link interference. Recent studies on dynamic time-division duplexing (TDD) have demonstrated that that UE-to-UE cross-link interference in distributed networks must be explicitly accounted for~\cite{Jay18, Zhang2019DynamicTDD,Liu2020DynamicTDD
}. Consequently, beamforming design for FD cell-free systems must incorporate UE-to-UE cross-link interference to fully realize the spectral efficiency gains of FD operation. Although OTA-based distributed beamforming has been successfully developed for half-duplex (HD) cell-free systems \cite{Atz21,Gouda2024CombinedDLUL,Gouda2024MulticastOTA}, its extension to FD operation introduces new challenges. In particular, simultaneous UL and DL pilot transmission results in UE-to-UE pilot leakage, which contaminates the OTA reconstruction of the cross-term channel information required for distributed beamforming design at the APs.

In this paper, we develop an OTA-based distributed beamforming framework for FD cell-free massive MIMO systems under a joint UL and DL sum mean-squared-error (MSE) criterion. Unlike prior HD OTA frameworks, FD operation introduces UE-to-UE interference and pilot leakage due to simultaneous UL and DL transmissions. We first derive the optimality conditions for the distributed UL and DL beamformers at the APs and UEs under perfect CSI, and iteratively update them within an alternating optimization framework. Building on this, the beamformers are reconstructed via IBT-based OTA signaling without explicit CSI exchange. As in prior OTA methods~\cite{Atz21,Gouda2024MulticastOTA}, cross terms for AP-specific updates are obtained via UE retransmissions; however, in FD systems, these signals are corrupted by UE-to-UE pilot leakage. To address this, we introduce a pilot-domain projection at the UEs prior to retransmission, which suppresses the interference while preserving the desired signal subspace, enabling accurate cross-term reconstruction. Furthermore, since UE beamformers jointly affect both UL and DL objectives, direct MSE-based updates may lead to unstable convergence under strong interference. We therefore employ best-response updates within the alternating optimization framework to improve convergence. Numerical results demonstrate faster convergence and higher effective sum rate, particularly for strongly interfering UEs, compared with separate UL/DL OTA training and local CSI-based designs.

\vspace{-3mm}
\section{System Model and Beamforming Design}
\label{sec:sys}

We consider a cell-free massive MIMO network with a set of distributed APs $\setB \triangleq \{1,\dots,B\}$, where each AP is equipped with $M$ antennas and connected to a central processing unit (CPU) via fronthaul links. The network serves a set of UEs $\setK \triangleq \{1,\dots,K\}$, each equipped with $N$ antennas. We assume a TDD system such that the UL and DL channels are reciprocal, and the UEs are partitioned into $\setK^{\dl}$ and $\setK^{\ul}$, containing $K_{\dl}$ and $K_{\ul}$ UEs, respectively. The UEs operate in HD mode, whereas all the APs operate in FD mode and can transmit and receive simultaneously over the same time--frequency resource. Let $\H_{b,k}\in\mathbb{C}^{M\times N}$ denote the uplink channel from UE $k$ to AP $b$. Similarly, the channel from DL UE $k\in\setK^{\dl}$ to  UL UE $u\in\setK^{\ul}$ is denoted by $\F_{k,u}\in\mathbb{C}^{N\times N}$, which captures the UE-to-UE interference arising from simultaneous UL and DL transmissions. 
Let $\S_{b, \bar b}\in\mathbb{C}^{M\times M}$ denote the coupling channel from AP $b$ to AP $\bar b$; for $b=\bar b$, this matrix models the residual self-interference at AP $b$ after analog cancellation.  In the following, we describe the DL and UL signal models under the FD cell-free system.

\smallskip
\noindent \textbf{Downlink signal model}. Let AP $b$ transmit the data symbol $d_k^{\dl}\sim\mathcal{CN}(0,1)$ to DL UE $k$ using the precoder $\w_{b,k}^{\dl}\in\mathbb{C}^{M\times1}$, while receiving the data symbol $d_u^{\ul}\sim\mathcal{CN}(0,1)$ from UL UE~$u$. Accordingly, the received signal at DL UE $k$ can be written as
\begin{align}
\y_k^{\dl} = \sum_{b\in\setB} \sum_{i\in\setK^{\dl}} \H_{b,k}^{\herm} \w_{b,i}^{\dl} d_i^{\dl} + \sum_{u\in\setK^{\ul}}
\F_{k,  u}^{\herm} \v_u^{\ul} d_u^{\ul} + \z_k^{\dl},
\end{align}
where $\v_u^{\ul}$ denotes the UL precoder of UE $u$, and $\z_k^{\dl}\sim\mathcal{CN}(\mathbf{0},\sigma_{\ue}^2\I_N)$ represents the additive white Gaussian noise (AWGN) at DL UE~$k$. The DL UE $k$ applies the receive combiner $\v_k^{\dl}\in\mathbb{C}^{N\times1}$ to obtain the soft estimate of $d_k^{\dl}$ as
\begin{align}
\hat d_k^{\dl} = (\v_k^{\dl})^{\herm}\y_k^{\dl}.
\end{align}
The resulting DL signal-to-interference and noise ratio (SINR) is given in~\eqref{eq:SINR_dl}.
\begin{figure*}
\begin{align}\label{eq:SINR_dl}
\gamma_k^{\dl}
=
\frac{
|\sum_{b \in \setB}(\v_k^{\dl})^{\herm}\H_{b,k}^{\herm}\w_{b,k}^{\dl}|^2
}{
\sum_{i\neq k}
|\sum_{b \in \setB}(\v_k^{\dl})^{\herm}\H_{b,k}^{\herm}\w_{b,i}^{\dl}|^2
+
\sum_{u\in\setK^{\ul}}
|(\v_k^{\dl})^{\herm}\F_{k,u}^{\herm}\v_u^{\ul}|^2
+
\sigma_{\ue}^2\|\v_k^{\dl}\|^2
}
\end{align}
\hrule
\vspace{-3mm}
\end{figure*}

\smallskip

\noindent \textbf{Uplink signal model.}
Let UL UE $u$ transmit the data symbol $d_u^{\ul}\sim\mathcal{CN}(0,1)$ using the precoder $\v_u^{\ul}\in\mathbb{C}^{N\times1}$. Accordingly, the signal received at AP $b$, while serving DL UEs is
\begin{align}
\y_b^{\ul}
=
\sum_{u\in\setK^{\ul}}
\H_{b,u}\v_u^{\ul} d_u^{\ul}
+
\sum_{\bar b\in\setB}
\sum_{i\in\setK^{\dl}}
\S_{b,\bar b}\w_{\bar b,i}^{\dl} d_i^{\dl}
+
\z_b^{\ul},
\end{align}
where $\z_b^{\ul}\sim\mathcal{CN}(\mathbf{0},\sigma_{\bs}^2\I_M)$ denotes the AWGN at AP~$b$.

Considering that APs know their transmitted DL symbols and precoders, they perform digital self-interference subtraction. Let 
$\hat{\g}_{b,i}$ denote the estimate of the leakage term $\sum_{\bar b\in\setB} \S_{b,\bar b}\w_{\bar b,i}^{\dl}$. The residual signal after subtraction becomes
\begin{align}
\tilde{\y}_b^{\ul}
=
\sum_{u\in\setK^{\ul}}
\H_{b,u}\v_u^{\ul} d_u^{\ul}
+
\sum_{i\in\setK^{\dl}}
\deltab_{b,i} d_i^{\dl}
+
\z_b^{\ul},
\end{align}
where $ {\deltab}_{b,i}
\triangleq \sum_{\bar b\in\setB} \S_{b,\bar b}\w_{\bar b,i}^{\dl} - \hat{\g}_{b,i}$ is the residual interference due to imperfect cancellation, which is modeled as an additive impairment dominated by the AWGN with mean $0$ and variance is a scaled identity matrix, and is therefore treated as statistically independent of the DL precoder. The AP $b$ applies the UL combiner $\w_{b,u}^{\ul}\in\mathbb{C}^{M\times1}$ and forwards the locally combined signal to CPU. The global soft estimate of  $d_u^{\ul}$ is
\begin{align}
\hat d_u^{\ul}
=
\sum_{b\in\setB}
(\w_{b,u}^{\ul})^{\herm}\tilde{\y}_b^{\ul}.
\end{align}
The resulting UL SINR is given in~\eqref{eq:SINR_ul}. Finally, the sum-rate of the UL and DL UEs is defined as
$R \triangleq \sum_{k\in \setK^{\dl}} \gamma_k^{\dl}
    + \sum_{u\in \setK^{\ul}} \gamma_u^{\ul},$ which serves as an upper bound on the achievable performance. In the following, we address the joint optimization of the DL and UL beamformers by minimizing the sum MSE of both DL and UL UEs under perfect CSI. This formulation serves as a reference benchmark for the pilot-based beamforming design discussed in Section~\ref{sec:ibt}.
\begin{figure*}
\begin{align}\label{eq:SINR_ul}
\gamma_u^{\ul}=\frac{\left|\sum_{b\in\setB}(\w_{b,u}^{\ul})^{\herm}\H_{b,u}\v_u^{\ul}\right|^2}{\sum_{j\neq u}\left|\sum_{b\in\setB}(\w_{b,u}^{\ul})^{\herm}\H_{b,j}\v_j^{\ul}\right|^2+\sum_{i\in\setK^{\dl}}\left|\sum_{b\in\setB}(\w_{b,u}^{\ul})^{\herm}\Delta_{b,i}\right|^2+\sigma_{\bs}^2\sum_{b\in\setB}\|\w_{b,u}^{\ul}\|^2}.
\end{align}
\hrule
\vspace{-3mm}
\end{figure*}

\subsection{Beamforming Design with Perfect CSI}
\label{sec:perfect}

Under perfect global CSI, we design DL precoders $\{\w_{b,k}^{\dl}\}$, DL combiners $\{\v_k^{\dl}\}$, UL precoders $\{\v_u^{\ul}\}$, and UL combiners $\{\w_{b,u}^{\ul}\}$ to minimize the  sum MSE of both DL and  UL UEs. Consequently, the MSE of DL UE $k\in\setK^{\dl}$ is defined as
\begin{align}
\mse_k^{\dl}\label{eq:mse_dl}
=
\E\!\left[
\left|
(\v_k^{\dl})^{\herm}\y_k^{\dl}
-
d_k^{\dl}
\right|^2
\right],
\end{align}
and the MSE of UL UE $u\in\setK^{\ul}$ is defined as,
\begin{align}
\mse_u^{\ul}\label{eq:mse_ul}
=
\E\!\bigg[
\Big|
\sum_{b\in\setB}
(\w_{b,u}^{\ul})^{\herm}\tilde{\y}_b^{\ul}
-
d_u^{\ul}
\Big|^2
\bigg].
\end{align}
Then, the joint beamforming design problem is formulated as
\begin{align} \label{eq:sum_mse_problem}
\begin{array}{cl}
\underset{{\{\w^{\dl},\v^{\dl},\w^{\ul},\v^{\ul}\}}}{\min}
&
\displaystyle \sum_{k\in\setK^{\dl}}\mse_k^{\dl}
+ \sum_{u\in\setK^{\ul}} \mse_u^{\ul} \\
\text{s.t.} &
\displaystyle  \sum_{k\in\setK^{\dl}}
\|\w_{b,k}^{\dl}\|^2 \le \rho_{\bs},\ \forall b,\\
& \|\v_u^{\ul}\|^2 \le \rho_{\ue},\ \forall u.
\end{array}
\end{align}
The above problem is non-convex with respect to the DL and UL beamformers. Therefore, we employ an alternating optimization approach in which each set of beamformers is updated while the remaining set of beamformers is kept fixed, as discussed in the following.

\smallskip
\noindent \textbf{DL UE combiner update}. For fixed a fixed set of $\{\w_{b,k}^{\dl}\}$, $\{\v_u^{\ul}\}$ , the combiner at DL UE~$k$ is obtained by minimizing the MSE in~\eqref{eq:mse_dl}. The resulting MMSE combiner is
\begin{align}
\v_k^{\dl} \!
=\!\!
\bigg(\!
\sum_{i\in\setK^{\dl}} {\h}^{\dl}_{k,i}({\h}^{\dl}_{k,i})^{\herm}
\!+\! \!
\sum_{u\in\setK^{\ul}}
{\f}^{\ul}_{k, u}  (\f^{\ul}_{k, u})^{\herm} \! \!+\!
\sigma_{\ue}^2\I
\bigg)^{-1}\!
\h_{k,k}^{\dl},
\label{eq:dl_mmse_v}
\end{align}
where $\h^{\dl}_{k,i} \triangleq \sum_{b\in\setB} \H_{b,k}^{\herm}\w_{b,i}^{\dl}$ and $\f^{\ul}_{k,u} \triangleq \F_{k,u}^{\herm}\v_u^{\ul}$.To improve convergence under strong UE–UE cross-link interference, we employ a best-response update within the alternating optimization framework, where each variable is optimized with the others fixed, leading to improved convergence behavior~\cite{Raz13}.

\smallskip
\noindent \textbf{AP-specific UL UE combiner update.}
For a fixed set of $\{\v_u^{\ul}\}$, AP $b$ updates $\w_{b,u}^{\ul}$ by minimizing $\mse_u^{\ul}$ in~\eqref{eq:mse_ul} while keeping $\{\w_{\bar b,u}^{\ul}\}_{\bar b\neq b}$ fixed from the previous iteration. 
The resulting AP-specific MMSE combiner for UL UE $u$ at AP $b$ is
\begin{align}
\w_{b,u}^{\ul}
=
\Big(
\Phi_{bb}^{\ul}
+
\mathbf{\Xi}_{b}^{\ul}
+
\nu_b\I + \sigma_{\bs}^2\I
\Big)^{-1}
\big(
\h_{b,u}^{\ul}
-
\xi_{b,u}^{\ul}
\big),
\label{eq:w_ul_star}
\end{align}
where
$\h_{b,j}^{\ul}
\triangleq
\H_{b,j}\v_j^{\ul}
$, $
\Phi_{b\bar b}^{\ul}
\triangleq
\sum_{j\in\setK^{\ul}}
\h_{b,j}^{\ul}
(\h_{\bar b,j}^{\ul})^{\herm}
$, $\mathbf{\Xi}_{b}^{\ul}
\triangleq
\sum_{i\in\setK^{\dl}}
\Delta_{b,i}
\Delta_{b,i}^{\herm}
$, and the cross term is $
\xi_{b,u}^{\ul}
\triangleq
\sum_{\bar b\in\setB\setminus\{b\}}
\Phi_{b\bar b}^{\ul}\,
\w_{\bar b,u}^{\ul}$.
The term $\nu_b\mathbf{I}$ denotes a regularization factor that improves convergence when appropriately chosen. The combiner is then updated using the best-response method to ensure stable convergence of the distributed MSE minimization~\cite{Gouda2021DistributedRx}. Note that the DL UE combiner and UL UE precoder updates are independent and can therefore be performed simultaneously.

\smallskip
\noindent \textbf{AP-specific DL UE precoder update.}
For a fixed set of $\{\v_k^{\dl}\}$, AP $b$ updates the DL precoder $\w_{b,k}^{\dl}$ by minimizing~\eqref{eq:sum_mse_problem}, while keeping $\{\w_{\bar b,k}^{\dl}\}_{\bar b\neq b}$ fixed from the previous iteration. The resulting MMSE precoder for DL UE $k$ at AP~$b$ is
\begin{align}
\w_{b,k}^{\dl}
=
\big(
\Phi_{bb}^{\dl}
+
\lambda_b \I
\big)^{-1}
\big(
\check{\h}_{b,k}^{\dl}
-
\xi_{b,k}^{\dl}
\big),
\label{eq:dl_precoder_single}
\end{align}
where $\check{\h}_{b,k}^{\dl}\triangleq \H_{b,k}\v_k^{\dl}$,
$\Phi_{b\bar b}^{\dl}
\triangleq
\sum_{i\in\setK^{\dl}}
\check{\h}_{b,i}^{\dl}(\check{\h}_{\bar b,i}^{\dl})^{\herm}$, and the cross term is
$\xi_{b,k}^{\dl}
\triangleq
\sum_{\bar b\in\setB\setminus\{b\}}
\Phi_{b\bar b}^{\dl}\,
\w_{\bar b,k}^{\dl}$. The scalar $\lambda_b\ge 0$ is chosen (e.g., via bisection) to satisfy the AP power constraint, with equality when active. The precoder is then updated using the best-response method to ensure convergence of the distributed MSE minimization~\cite{Atz21}.

\smallskip
\noindent \textbf{UL UE precoder update}. 
For a fixed set of $\{\w_{b,u}^{\ul}\}$,  $\{\v_k^{\dl}\}$, the precoder at UL UE~$u$ is obtained by minimizing~\eqref{eq:sum_mse_problem}. The resulting MMSE precoder is
\begin{align}
\v_u^{\ul}
=
\bigg( \!
\sum_{j\in\setK^{\ul}}
\check{\h}_{u,j}^{\ul}(\check{\h}_{u,j}^{\ul})^{\herm}
\!+\!\!
\sum_{k\in\setK^{\dl}}
{\f}_{k,u}^{\dl}({\f}_{k, u}^{\dl})^{\herm}
\!+\!
\mu_u\I
\bigg)^{-1}
\check{\h}_{u,u}^{\ul},
\label{eq:ul_precoder_mmse}
\end{align}
where $\check{\h}_{u,j}^{\ul}
\triangleq
\sum_{b\in\setB}
\H_{b,j}^{\herm}\w_{b,u}^{\ul}
$, ${\f}_{k, u}^{\dl}
\triangleq
\F_{k,u}\v_k^{\dl}
$,
and $\mu_u\ge 0$ is chosen (e.g., via bisection) to satisfy the UE power constraint, with equality when active. Similar to the DL combiner update, to improve convergence under strong UE–UE cross-link interference, we employ the regularized update for the UL UE precoder, where the step size is chosen based on the relative UE–UE interference power compared to the overall received signal power~\cite{Raz13}. Moreover, the DL UE combiner and UL UE precoder updates are independent and can therefore be performed simultaneously.
\vspace{-3mm}
\section{Beamforming Design in FD via OTA Training}
\label{sec:ibt}

This section presents a practical distributed implementation of the beamformers discussed in Section~\ref{sec:perfect} using OTA IBT. To this end, we assign orthogonal pilot sequences $\p_k \in \mathbb{C}^{\tau\times 1}$ and $\q_u \in \mathbb{C}^{\tau\times 1}$ to DL UEs $k \in \setK_{\dl}$ and UL UEs $u \in \setK_{\ul}$, respectively, and define $\P \triangleq [\p_1,\ldots,\p_{K_{\dl}}]$ and $\Q \triangleq [\q_1,\ldots,\q_{K_{\ul}}]$. The pilots satisfy $[\P, \Q]^{\herm}[\P, \Q] = \tau \I$, requiring $\tau \ge K_{\dl} + K_{\ul}$ and ensuring separation of UL and DL pilot subspaces, thereby preventing spurious cross terms in OTA beamforming design.

In the first time slot, all the APs transmit precoded DL pilots and simultaneously all the UL UEs transmit precoded UL pilots. Specifically, AP $b$ transmits
\begin{align}
\X_{b}^{\dlA} \triangleq \sum_{k\in\setK^{\dl}} \w_{b,k}^{\dl}\p_k^{\herm}\in\mathbb{C}^{M\times \tau},
\end{align}
and UL UE $u$ transmits
\begin{align}
\X_{u}^{\ulA} \triangleq \v_u^{\ul}\q_u^{\herm}\in\mathbb{C}^{N\times \tau}.
\end{align}
Accordingly, the received pilot signal at AP $b$ is
\begin{align}
\Y_{b}^{\ulA}
&=
\sum_{u\in\setK^{\ul}}\H_{b,u}\X_{u}^{\ulA} 
+
\sum_{\bar b\in\setB}
\S_{b,\bar b}\X_{b}^{\dlA}
+\Z_{b}^{\ulA}.
\label{eq:Yb1}
\end{align}
where $\Z_{b}^{\ulA}$ is the AWGN at AP~$b$, and each DL UE $k$ receives
\begin{align}
\Y_{k}^{\dlA}
&=
\sum_{i\in\setK^{\dl}}\H_{k}^{\herm}\X_{b}^{\dlA}
+
\sum_{u\in\setK^{\ul}}\F_{k,u}^{\herm}\X_{u}^{\ulA}
+\Z_{k}^{\dlA}.
\label{eq:Yk1}
\end{align}
where $\Z_{k}^{\dlA}$ is the AWGN at UE~$k$.

\smallskip
\noindent \textbf{DL UE combiner update.} From the received $\Y_{k}^{\dlA}$ in \eqref{eq:Yk1}, UE~$k$ obtains $\v_k^{\dl}$ as
\begin{align}\label{eq:v_dl_ota}
    \v_k^{\dl} = \big(\Y_{k}^{\dlA}(\Y_{k}^{\dlA})^{\herm}\big)^{-1}\Y_{k}^{\dlA}\p_k,
\end{align}
which converges to~\eqref{eq:dl_mmse_v} as $\tau \rightarrow \infty$ or at high SINR.

In the second time slot, each AP~$b$ transmits pilots to UL UEs after precoding with the respective combiners as
\begin{align}
\X_{b}^{\dlB} \triangleq \sum_{u\in\setK^{\ul}} \w_{b,u}^{\ul}\q_u^{\herm}/{\sqrt{\beta_1}} \ \in\mathbb{C}^{M\times \tau},
\end{align}
where $\beta_1$ is a scaling factor to meet the power constraint at APs. Simultaneously, each DL UE $k$ transmits the pilots after precoding with its combiner as
\begin{align}
\X_{k}^{\ulB} \triangleq  \v_k^{\dl}\p_k^{\herm}/ {\sqrt{\beta_2}} \ \in\mathbb{C}^{N\times \tau},
\end{align}
where $\beta_2$ is the scaling factor to meet the UE power constraint. Then, the corresponding received pilot signal at AP $b$ is
\begin{align}
\Y_{b}^{\ulB}
&=
\sum_{k\in\setK^{\dl}}\H_{b,k}\X_{k}^{\ulB}
+
\sum_{\bar b\in\setB}\S_{b,\bar b}\X_{\bar b}^{\dlB}
+
\Z_{b}^{\ulB},
\label{eq:Y_b_B}
\end{align}
where $\Z^{\ulB}_{b}$ is the AWGN at AP $b$, and each UL UE $u$ receives
\begin{align}
\Y_{u}^{\dlB}
&=
\sum_{b\in\setB}\H_{b,u}^{\herm}\X_{b}^{\dlB}
+
\sum_{k\in\setK^{\dl}}\F_{k,u}\X_{k}^{\ulB}
+
\Z_{u}^{\dlB},
\label{eq:Y_u_B}
\end{align}
where $\Z_{u}^{\dlB}$ denotes the AWGN at UL UE $u$.

\smallskip
\noindent \textbf{UL UE precoder update.}
From $\Y_{u}^{\dlB}$ in~\eqref{eq:Y_u_B}, UL UE $u$ updates $\v_u^{\ul}$ as
\begin{align}
\v_u^{\ul}
=
\big(\Y_{u}^{\dlB}(\Y_{u}^{\dlB})^{\herm} + \tau \mu_u \I \big)^{-1}\Y_{u}^{\dlB}\q_u,
\label{eq:v_ul_ota}
\end{align}
where $\mu_u\ge 0$ is obtained via bisection to satisfy the UE power constraint. The update in~\eqref{eq:v_ul_ota} converges to~\eqref{eq:ul_precoder_mmse} as $\tau\rightarrow\infty$ or at high SINR.

In contrast to conventional HD OTA beamforming training frameworks \cite{Atz21,Gouda2024MulticastOTA,Gouda2024CombinedDLUL}, where an additional uplink signaling phase is used to reconstruct the cross terms, such as $\xi_{b,k}^{\dl}$ and $\xi_{b,u}^{\ul}$, at each AP without backhaul signaling, the simultaneous transmission of UL and DL pilots in FD operation introduces UE-to-UE pilot leakage into the received signals at the UEs. As a result, the conventional cross-term reconstruction mechanism is no longer directly applicable. To address this, in the following, we propose a projection-based modification for the extra uplink signaling, which is essential in FD operation to suppress the undesired UE-to-UE interference component while preserving the terms required for the AP-specific beamformer updates. To this end, to update the AP-specific DL precoders in~\eqref{eq:dl_precoder_single} and UL combiners in~\eqref{eq:w_ul_star}, each DL UE retransmits $\Y_{k}^{\dlA}$ in~\eqref{eq:Yk1} during the third time slot after projection onto the subspace of $\P$ and precoding with $\v_k^{\dl}(\v_k^{\dl})^{\herm}$, given by
\begin{align}
\X^{\ulC-\dl}_{k}
\triangleq
\v_k^{\dl}(\v_k^{\dl})^{\herm}
\Y_{k}^{\dlA}\P \P^{\herm}/({\tau \sqrt{\beta_2}}).
\end{align}
The projection of $\Y_{k}^{\dlA}$ onto the subspace of $\P$ removes the UE–UE interference component $\sum_{u\in\setK^{\ul}}\F_{k,u}^{\herm}\X_{u}^{\ulA}$ and enables the APs to reconstruct the cross terms $\xi_{b,k}^{\dl}$ required for the DL precoder update in~\eqref{eq:dl_precoder_single}.

Similarly, during the third time instant, each UL UE also retransmits the received signal $\Y_{u}^{\dlB}$ in~\eqref{eq:Y_u_B} after projecting onto the subspace of $\Q$ and precoding with $\v_u^{\ul}(\v_u^{\ul})^{\herm}$ as
\begin{align}
\X^{\ulC-\ul}_{u}
\triangleq
\sqrt{\beta_1}
\v_u^{\ul}(\v_u^{\ul})^{\herm}
\Y_{u}^{\dlB}\Q \Q^{\herm}/\tau .
\end{align}
The projection of $\Y_{u}^{\dlB}$ onto $\Q$ removes the UE–UE interference term $\sum_{k\in\setK^{\dl}}\F_{k,u}\X_{k}^{\ulB}$ and enables the APs to reconstruct the cross terms $\xi_{b,u}^{\ul}$ required for the UL combiner update in~\eqref{eq:w_ul_star}. Accordingly, the received signal at AP $b$ is
\begin{align}
\Y_{b}^{\ulC} \!
&= \!
\sum_{k\in\setK^{\dl}}
\H_{b,k}\X^{\ulC-\dl}_{k}
\!+\!
\sum_{u\in\setK^{\ul}}
\H_{b,u}\X^{\ulC-\ul}_{u}
\!+\!
\Z_{b}^{\ulC},
\label{eq:YbC}
\end{align}
where $\Z_{b}^{\ulC}$ is the AWGN at AP~$b$.

\smallskip
\noindent \textbf{AP-specific DL UE precoder update.}
Using the local received pilot $\Y_{b}^{\ulB}$ and $\Y_{b}^{\ulC}$, AP $b$ updates the DL precoder for DL UE $k$ as in~\eqref{eq:w_dl_ota}.
\begin{figure*}
\begin{align}
\w_{b,k}^{\dl}
&=
\Big(
\Y_{b}^{\ulB}\P \P^{\herm}(\Y_{b}^{\ulB})^{\herm}
+
\beta_2\tau^2 (\lambda_b-\sigma^2_{\ue})\I
\Big)^{-1}
\Big(
\tau \sqrt{\beta_2}\Y_{b}^{\ulB}\p_k
-
\Big(
{\tau} \sqrt{\beta_2}\Y_{b}^{\ulC}\p_k
-
\Y_{b}^{\ulB}\P \P^{\herm}(\Y_{b}^{\ulB})^{\herm}\w_{b,k}^{\dl,{\rm old}}
\Big)
\Big)
\label{eq:w_dl_ota}
\end{align}
\hrule
\vspace{-3mm}
\end{figure*}
Here $\w_{b,k}^{\dl,{\rm old}}$ is the precoder used in the  previous iteration, and $\lambda_b\ge 0$ is computed via bisection to satisfy the power constraint of AP. The precoder update in~\eqref{eq:w_dl_ota} converges to~\eqref{eq:dl_precoder_single} as $\tau\rightarrow\infty$ or at high SINR.

\smallskip
\noindent \textbf{AP-specific UL UE combiner update.}
Similarly, using $\Y_{b}^{\ulA}$ and $\Y_{b}^{\ulC}$, AP $b$ updates the UL combiner for UL UE $u$ as in~\eqref{eq:w_ul_ota}.
\begin{figure*}
\begin{align}
\w_{b,u}^{\ul}
&=
\Big(
{\Y}_{b}^{\ulA}\Q\Q^{\herm}({\Y}_{b}^{\ulA})^{\herm}
+
\tau^2 \bar \nu_{b}\I
\Big)^{-1}
\Big(
{\tau}{\Y}_{b}^{\ulA}\q_u
-
\Big(
{\tau}\Y_{b}^{\ulC}\q_u
-
{\Y}_{b}^{\ulA} \Q \Q^{\herm}({\Y}_{b}^{\ulA})^{\herm}\w_{b,u}^{\ul,{\rm old}}
\Big)
\Big)
\label{eq:w_ul_ota}
\end{align}
\hrule
\vspace{-3mm}
\end{figure*}
Here $\bar{\nu}{b}$ denotes the effective variance accounting for self-interference, AWGN, and the regularization term discussed in Section~\ref{sec:perfect}, and $\w{b,u}^{\ul,{\rm old}}$ denotes the combiner used in the previous iteration. The combiner update in~\eqref{eq:w_ul_ota} converges to~\eqref{eq:w_ul_star} as $\tau\rightarrow\infty$ or at high SINR. Finally, all the beamformers are updated using the best-response method as discussed in Section~\ref{sec:perfect}. The complete procedure to obtain DL and UL beamformers using IBT framework is given in Algorithm~\ref{alg:fd_ota}.

\begin{figure}[t]
\begin{algorithm}[H]
\small
\begin{spacing}{1.2}
\textbf{Data:} DL UE pilots $\{\p_k\}_{k\in\setK^{\dl}}$, and UL UE pilots $\{\q_u\}_{u\in\setK^{\ul}}$.\\
\textbf{Initialization:} $\{\w_{b,k}^{\dl}\}$, $\{\v_u^{\ul}\}$, and scaling factors $\beta_1,\beta_2$.\\
\textbf{For each IBT iteration}, \textbf{do:}
\begin{itemize}[leftmargin=*]
\item[1)]  
\textbf{Slot-1:} Each AP $b$ transmits $\X_{b}^{\dlA}$ and each UL UE $u$ transmits $\X_{u}^{\ulA}$. 
Each AP~$b$ receives $\Y_{b}^{\ulA}$ in~\eqref{eq:Yb1} and each DL UE $k$ receives $\Y_{k}^{\dlA}$ in~\eqref{eq:Yk1}. 
\item[2)]
Each DL UE $k$ computes $\v_k^{\dl}$ from $\Y_{k}^{\dlA}$ as in~\eqref{eq:v_dl_ota}.
\item[3)]
\textbf{Slot-2:} Each AP $b$ transmits $\X_{b}^{\dlB}$ and each DL UE $k$ transmits $\X_{k}^{\ulB}$. 
Each AP receives $\Y_{b}^{\ulB}$ in~\eqref{eq:Y_b_B} and each UL UE $u$ receives $\Y_{u}^{\dlB}$ in~\eqref{eq:Y_u_B}.
\item[4)]
Each UL UE $u$ computes $\v_u^{\ul}$ from $\Y_{u}^{\dlB}$ as in~\eqref{eq:v_ul_ota}.
\item[5)] 
\textbf{Slot-3:} Each DL UE $k$ transmits $\X_{k}^{\ulC-\dl}$ and each UL UE $u$ transmits $\X_{u}^{\ulC-\ul}$. 
Each AP $b$ receives $\Y_{b}^{\ulC}$ as in~\eqref{eq:YbC}.
\item[6)] Each AP $b$ updates $\w_{b,k}^{\dl}$ using $\Y_{b}^{\ulB}$ and $\Y_{b}^{\ulC}$ as in~\eqref{eq:w_dl_ota}.
\item[7)] Each AP $b$ updates $\w_{b,u}^{\ul}$ using ${\Y}_{b}^{\ulA}$ and $\Y_{b}^{\ulC}$ as in~\eqref{eq:w_ul_ota}.
\end{itemize}
\textbf{Output:} $\{\w_{b,k}^{\dl}\}$, $\{\v_k^{\dl}\}$, $\{\v_u^{\ul}\}$, and $\{\w_{b,u}^{\ul}\}$.
\end{spacing}
\caption{(FD-OTA distributed beamforming design)} \label{alg:fd_ota}
\end{algorithm}
\vspace{-6mm}
\end{figure}

\section{Numerical Results and Discussion}\label{sec:num}

We consider an FD cell-free network with $B=16$ APs, each equipped with $M=4$ antennas and deployed on a square grid with an inter-site distance of $100$~m. A total of $K_{\dl}=16$ DL UEs and $K_{\ul}=16$ UL UEs are uniformly distributed in the coverage area, each equipped with $N=4$ antennas. The AP-to-UE channel $\H_{b,k}$ is generated as $\mathrm{vec}(\H_{b,k}) \sim \mathcal{CN}(0,\delta_{b,k}\I_{MN})$, where the large-scale fading follows the pathloss model $\delta_{b,k}\,[\mathrm{dB}] = -30.5 - 37\log_{10}(d_{b,k})$ for a $2.5$~GHz carrier frequency. The UL-to-DL interference channel $\F_{k,u}$ is generated similarly with an additional $-20$~dB isolation between UEs. The AP-to-AP coupling matrix $\S_{\bar b, b}$ follows the same model for $\bar b \neq b$, while for $\bar b=b$ an additional $-40$~dB attenuation is applied to represent residual self-interference. The maximum transmit power at AP and UE is $\rho_{\bs}=\rho_{\ue}=30$~dBm, and the AWGN power at both APs and UEs is fixed at $-95$~dBm. The UL and DL pilot length is $\tau=32$. The self-interference term $\g_{b,i}$ is estimated at AP $b$ as $\hat{\g}_{b,i}=\frac{1}{\tau}\Y_{b}^{\ulA}\q_i $, which is then used for self-interference cancellation during UL reception, and the performance is evaluated using the sum-rate $R$, given in Section~\ref{sec:sys}.

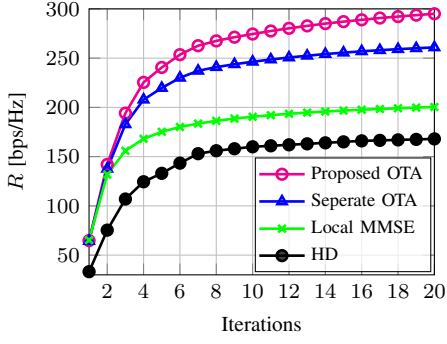
\begin{figure*}
    \begin{minipage}[t]{0.3\textwidth}
        \begin{figure}[H]
            \input{figures/fig1.tex}
            \vspace{-4mm}
\caption{\small{Sum of UL and DL rates versus IBT iterations.}}
\label{fig:rateVsItr}
        \end{figure}
    \end{minipage}
    \hfill
    \begin{minipage}[t]{0.3\textwidth}
        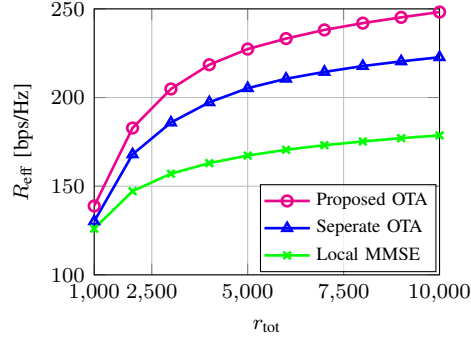
\begin{figure}[H]
            \input{figures/fig3.tex}
            \vspace{-4mm}
\caption{\small{Sum of effective UL and DL rates versus available resources.}}
\label{fig:effrateVsItr}
        \end{figure}
    \end{minipage}
    \hfill
    \begin{minipage}[t]{0.3\textwidth}
        \begin{figure}[H]
            \input{figures/fig2.tex}
            \vspace{-4mm}
\caption{\small{CDF of UE rates (solid: DL UE rate, dashed: UL UE rate).}}
\label{fig:rateVcdf}
        \end{figure}
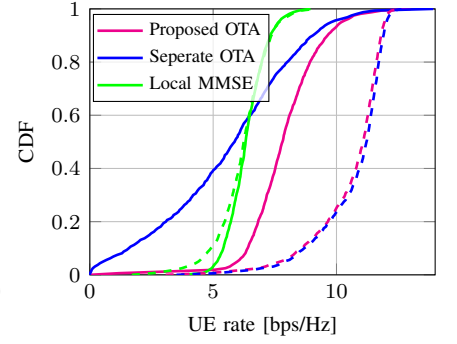
    \end{minipage}
    \vspace{-5mm}
\end{figure*}



In Fig.~\ref{fig:rateVsItr}, we plot the sum of UL and DL rates versus the IBT iterations. The proposed OTA scheme is compared with \emph{Separate OTA training}, where the DL and UL beamformers are trained independently and thus do not account for UE-to-UE cross-link interference, and with the \emph{Local MMSE} baseline, which does not exploit the UL-3 OTA signaling and therefore neglects the cross terms required in the AP-specific UL combiner and DL precoder updates. By explicitly accounting for UE-to-UE interference and reconstructing the required cross terms via the UL-3 OTA step, the proposed method achieves a higher sum of the UL and DL rates than both baseline schemes. Finally, we also compare with an HD scenario, where beamformer training for DL and UL UEs is performed separately, and the APs serve DL and UL UEs on orthogonal resources with equal allocation, which is spectrally inefficient compared to FD operation.

In Fig.~\ref{fig:effrateVsItr}, we illustrate the effective sum-rate performance of all methods as a function of the available resources. The effective rate is defined as $ R_{\text{eff}} \triangleq 
\left(1 - {t\, r_{\text{IBT}}}/{r_{\text{tot}}}\right) R,$
where $t$ denotes the number of iterations, $r_{\text{IBT}}$ is the resource consumption per IBT iteration, and $r_{\text{tot}}$ is the total available resource for training and data transmission. The separate and proposed methods require $3\tau$ resources per iteration, whereas the local MMSE scheme requires $2\tau$ resources per iteration. The results show that even for small $r_{\text{tot}}$, the proposed method outperforms both the separate OTA and local MMSE schemes. Moreover, as $r_{\text{tot}}$ increases, the relative training overhead becomes negligible and the performance gap further widens.

In Fig.~\ref{fig:rateVcdf}, we plot the CDF of the per-UE rates at iteration~20. Since the proposed method explicitly accounts for UE--UE cross-link interference in the UL precoder design, the UL UE rates may be slightly reduced compared to the Separate OTA training scheme. However, the interference-aware UL precoder together with the coupled DL combiner updates substantially improves the DL UE rates of the proposed method, particularly for strongly interfering UEs.

\section{Conclusion}
We developed an OTA-based distributed beamforming framework for FD cell-free massive MIMO systems under a joint UL and DL sum-MSE criterion. To mitigate the UE-to-UE pilot leakage caused by simultaneous UL and DL pilot transmission, we introduced a pilot-domain projection at the UEs prior to retransmission, enabling accurate cross-term reconstruction for AP-specific beamformer updates. In addition, regularized best-response updates at the UEs improve convergence robustness under strong UE-to-UE interference. Numerical results demonstrated faster convergence and higher effective sum rate than separate UL and DL OTA training and local CSI-based beamforming designs, with particularly larger gains for strongly interfering UEs.

\bibliographystyle{IEEEtran}
\bibliography{IEEEabbr,ref2}

\end{document}

%% file: Definitions.tex
\newcommand{\f}{\mathbf{f}}
\newcommand{\g}{\mathbf{g}}
\newcommand{\h}{\mathbf{h}}

\newcommand{\p}{\mathbf{p}}
\newcommand{\q}{\mathbf{q}}

\renewcommand{\v}{\mathbf{v}}
\newcommand{\w}{\mathbf{w}}

\newcommand{\y}{\mathbf{y}}
\newcommand{\z}{\mathbf{z}}

\newcommand{\E}{\mathbf{E}}
\newcommand{\F}{\mathbf{F}}

\renewcommand{\H}{\mathbf{H}}
\newcommand{\I}{\mathbf{I}}

\renewcommand{\P}{\mathbf{P}}
\newcommand{\Q}{\mathbf{Q}}

\renewcommand{\S}{\mathbf{S}}

\newcommand{\X}{\mathbf{X}}
\newcommand{\Y}{\mathbf{Y}}
\newcommand{\Z}{\mathbf{Z}}

\newcommand{\deltab}{\boldsymbol{\delta}}

\newcommand{\setB}{\mathcal{B}}

\newcommand{\setK}{\mathcal{K}}

\newcommand{\herm}{\mathrm{H}}



%% file: figures/fig1.tex
\begin{tikzpicture}

\begin{axis}[
	width=6.15cm,
	height=5.1cm,
	xmin=1, xmax=20,
	ymin=30, ymax=300,
    xlabel={Iterations},
    ylabel={$R$ [bps/Hz]},
	ytick={50,100,150,200,250,300},
    xtick={2,4,6,8,10,12,14,16,18,20},
    xlabel near ticks,
	ylabel near ticks,
    x label style={font=\footnotesize},
	y label style={font=\footnotesize},
    ticklabel style={font=\footnotesize},
    legend pos=south west,
    legend cell align=left,
    legend style={at={(0.48,0.01)}, anchor=south west},
	legend style={font=\scriptsize, inner sep=1pt, fill opacity=0.75, draw opacity=1, text opacity=1},
	grid=both,
]

\addplot[line width=1pt, magenta, mark=o, mark options={solid}]
table[x=Itr, y=Proposed, col sep=comma] 
{figures/fig1_data/data.txt};
\addlegendentry{{Proposed OTA}};

\addplot[line width=1pt, blue, mark=triangle, mark options={solid}]
table[x=Itr, y =Seperate, col sep=comma] 
{figures/fig1_data/data.txt};
\addlegendentry{{Seperate OTA}};

\addplot[line width=1pt,  green, mark=x, mark options={solid}]
table[x=Itr, y =Local, col sep=comma] 
{figures/fig1_data/data.txt};
\addlegendentry{{Local MMSE}};

\addplot[line width=1pt,  black, mark=*, mark options={solid}]
table[x=Itr, y =HD, col sep=comma] 
{figures/fig1_data/data.txt};
\addlegendentry{{HD}};

\end{axis}

\end{tikzpicture}

%% file: figures/fig3.tex
\begin{tikzpicture}

\begin{axis}[
	width=6.15cm,
	height=5.1cm,
	xmin=1000, xmax=10000,
	ymin=100, ymax=250,
    xlabel={$r_{\text{tot}}$},
    ylabel={$R_{\text{eff}}$ [bps/Hz]},
	ytick={100,150,200,250},
   xtick={1000,2500,5000,7500,10000},
    x label style={font=\footnotesize},
	y label style={font=\footnotesize},
    ticklabel style={font=\footnotesize},
    legend pos=south west,
    legend cell align=left,
    legend style={at={(0.48,0.01)}, anchor=south west},
	legend style={font=\scriptsize, inner sep=1pt, fill opacity=0.75, draw opacity=1, text opacity=1},
	grid=both,
	scaled x ticks= false,
]

\addplot[line width=1pt, magenta, mark=o, mark options={solid}]
table[x=Res, y=Proposed, col sep=comma] 
{figures/fig3_data/data.txt};
\addlegendentry{{Proposed OTA}};

\addplot[line width=1pt, blue, mark=triangle, mark options={solid}]
table[x=Res, y =Seperate, col sep=comma] 
{figures/fig3_data/data.txt};
\addlegendentry{{Seperate OTA}};

\addplot[line width=1pt,  green, mark=x, mark options={solid}]
table[x=Res, y =Local, col sep=comma] 
{figures/fig3_data/data.txt};
\addlegendentry{{Local MMSE}};

\end{axis}

\end{tikzpicture}

%% file: figures/fig2.tex
\begin{tikzpicture}

\begin{axis}[
	width=6.15cm,
	height=5.1cm,
	xmin=0, xmax=14,
	ymin=0, ymax=1,
    xlabel={UE rate [bps/Hz]},
    ylabel={CDF },
    xlabel near ticks,
	ylabel near ticks,
    x label style={font=\footnotesize},
	y label style={font=\footnotesize},
    ticklabel style={font=\footnotesize},
    legend pos=south east,
    legend cell align=left,
    legend style={at={(0.52,0.65)}, anchor=south east},
	legend style={font=\scriptsize, inner sep=1pt, fill opacity=0.75, draw opacity=1, text opacity=1},
	grid=both,
]

\addplot[line width=1pt, magenta, mark options={solid}]
table[x=y, y=x , col sep=comma] 
{figures/fig2_data/data_p_dl.txt};
\addlegendentry{{Proposed OTA}};

\addplot[line width=1pt, blue, mark options={solid}]
table[x=y, y=x , col sep=comma] 
{figures/fig2_data/data_b_dl.txt};
\addlegendentry{{Seperate OTA}};

\addplot[line width=1pt, green,  mark options={solid}]
table[x=y, y=x , col sep=comma] 
{figures/fig2_data/data_M_dl.txt};
\addlegendentry{{Local MMSE}};

\addplot[line width=1pt, magenta, dashed, mark options={solid}]
table[x=y, y=x , col sep=comma] 
{figures/fig2_data/data_p_ul.txt};

\addplot[line width=1pt, blue, dashed, mark options={solid}]
table[x=y, y =x , col sep=comma] 
{figures/fig2_data/data_b_ul.txt};

\addplot[line width=1pt, green, dashed, mark options={solid}]
table[x=y, y =x , col sep=comma] 
{figures/fig2_data/data_M_ul.txt};

\end{axis}

\end{tikzpicture}

%% file: IEEEabbr.bib
@STRING{J_IEEE_TC      		= "IEEE Trans. Commun."}

@STRING{J_IEEE_JSAC			= "IEEE J. Sel. Areas Commun."}

@STRING{J_IEEE_TSP			= "IEEE Trans. Signal Process."}

@STRING{J_IEEE_TWC			= "IEEE Trans. Wireless Commun."}

@STRING{J_SPRINGER_EJWCN	= "EURASIP J. Wireless Commun. Netw."}

@STRING{L_IEEE_WCL			= "IEEE Wireless Commun. Lett."}

@STRING{C_IEEE_ASILOMAR		= "Proc. Asilomar Conf. Signals, Syst., and Comput. (ASILOMAR)"}

@STRING{C_IEEE_ICC			= "Proc. IEEE Int. Conf. Commun. (ICC)"}


%% file: ref2.bib
@article{Ngo2017CellFree,
  author  = {H. Q. Ngo and A. Ashikhmin and H. Yang and E. G. Larsson and T. L. Marzetta},
  title   = {Cell-Free Massive {MIMO} Versus Small Cells},
  journal = J_IEEE_TWC,
  year    = {2017},
  volume  = {16},
  number  = {3},
  pages   = {1834--1850},
  month   = mar
}

@inproceedings{Nayebi2017PerformanceCF,
  author    = {E. Nayebi and A. Ashikhmin and T. L. Marzetta and H. Yang and B. D. Rao},
  title     = {Performance of Cell-Free Massive {MIMO} Systems with {MMSE} and {LSFD} Receivers},
  booktitle = C_IEEE_ASILOMAR,
  year      = {2017},
  pages    = {203--207},
  month = nov
}

@article{Interdonato2019ScalableCF,
  author  = {G. Interdonato and E. Bj{\"o}rnson and H. Q. Ngo and P. Frenger and E. G. Larsson},
  title   = {Ubiquitous Cell-Free Massive {MIMO} Communications},
  journal = J_SPRINGER_EJWCN,
  year    = {2019},
  volume  = {2019},
  number  = {197},
  month   = aug
}

@article{Bjornson2019MakingCFScalable,
  author  = {E. Bj{\"o}rnson and L. Sanguinetti},
  title   = {Making Cell-Free Massive {MIMO} Competitive With {MMSE} Processing and Centralized Implementation},
  journal = J_IEEE_TWC,
  year    = {2020},
  volume  = {19},
  number  = {1},
  pages   = {77--90},
  month   = jan
}

@article{Rogalin2014ScalableBF,
  author  = {R. Rogalin and O. Y. Bursalioglu and H. Papadopoulos and G. Caire and A. F. Molisch and A. Michaloliakos and V. Balan and K. Psounis},
  title   = {Scalable Synchronization and Reciprocity Calibration for Distributed Multiuser {MIMO}},
  journal = J_IEEE_TWC,
  year    = {2014},
  volume  = {13},
  number  = {4},
  pages   = {1815--1831},
  month   = apr,
  doi     = {10.1109/TWC.2014.030314.130474}
}

@article{Shi2011WMMSE,
  author  = {Q. Shi and M. Razaviyayn and Z.-Q. Luo and C. He},
  title   = {An Iteratively Weighted {MMSE} Approach to Distributed Sum-Utility Maximization for a {MIMO} Interfering Broadcast Channel},
  journal = J_IEEE_TSP,
  year    = {2011},
  volume  = {59},
  number  = {9},
  pages   = {4331--4340},
  month   = sep
}

@article{Gouda2024MulticastOTA,
  author  = {B. Gouda and I. Atzeni and A. T{\"o}lli},
  title   = {Pilot-Aided Distributed Multi-Group Multicast Precoding Design for Cell-Free Massive {MIMO}},
  journal = J_IEEE_TWC,
  year    = {2024},
 volume={23},
  number={8},
  pages={9282--9298},
  month   = aug
}

@inproceedings{Gouda2021DistributedRx,
  author    = {B. Gouda and A. T{\"o}lli},
  title     = {Distributed Joint Receiver Design for Cell-Free Massive {MIMO} with Fast Convergence},
  booktitle = C_IEEE_ICC,
  year      = {2021},
   volume={},
  number={},
  pages={1--6},
  month = jun

  
}

@article{Sabharwal2014FDReview,
  author  = {A. Sabharwal and P. Schniter and D. Guo and D. W. Bliss and S. Rangarajan and R. Wichman},
  title   = {In-Band Full-Duplex Wireless: Challenges and Opportunities},
  journal = J_IEEE_JSAC,
  year    = {2014},
  volume  = {32},
  number  = {9},
  pages   = {1637--1652},
  month   = sep,
  doi     = {10.1109/JSAC.2014.2330193}
}

@inproceedings{Bharadia2013FD, 
author = {Bharadia, Dinesh and McMilin, Emily and Katti, Sachin}, title = {Full duplex radios}, 
year = {2013}, 
isbn = {9781450320566}, 
publisher = {Association for Computing Machinery},
address = {New York, NY, USA}, 
doi = {10.1145/2486001.2486033}, 
booktitle = {Proc. ACM SIGCOMM},
pages = {375--386}, 
numpages = {12}, 
location = {Hong Kong, China}, 
series = {SIGCOMM '13} }

@article{Everett2014PassiveSI,
  author  = {E. Everett and A. Sahai and A. Sabharwal},
  title   = {Passive Self-Interference Suppression for Full-Duplex Infrastructure Nodes},
  journal = J_IEEE_TWC,
  year    = {2014},
  volume  = {13},
  number  = {2},
  pages   = {680--694},
  month   = feb
}

@article{Zhang2019DynamicTDD,
  author  = {J. Zhang and E. Bj{\"o}rnson and M. Matthaiou and D. J. Love and D. W. K. Ng and H. Yang},
  title   = {Prospective Multiple Antenna Technologies for Beyond {5G}},
  journal = J_IEEE_JSAC,
  year    = {2020},
  volume  = {38},
  number  = {8},
  pages   = {1637--1660},
  month   = aug
}

@ARTICLE{Liu2020DynamicTDD,
  author={Chowdhury, Anubhab and Chopra, Ribhu and Murthy, Chandra R.},
  journal=J_IEEE_TC, 
  title={Can Dynamic {TDD} Enabled Half-Duplex Cell-Free Massive {MIMO} Outperform Full-Duplex Cellular Massive {MIMO}?}, 
  year={2022},
  volume={70},
  number={7},
  pages={4867--4883},
  month=Jul
 }

@article{Gouda2024CombinedDLUL,
  author  = {B. Gouda and A. Arvola and I. Atzeni and A. T{\"o}lli},
  title   = {Combined {DL}-{UL} Distributed Beamforming Design for Cell-Free Massive {MIMO}},
  journal = L_IEEE_WCL,
  year    = {2024},
  volume  = {13},
  number  = {6},
  pages   = {1621--1625},
  month   = jun
}

@article{Atz21,
	author = {I. {Atzeni} and B. {Gouda} and A. {Tölli}},
	title = {Distributed precoding design via over-the-air signaling for cell-free massive {MIMO}},
	journal = J_IEEE_TWC,
	volume = {20},
	number = {2},
	pages = {1201--1216},
	year = {2021},
    month=feb
}

@article{Jay18,
	author = {Jayasinghe, P. and T\"{o}lli, A. and Kaleva, J. and Latva-aho, M.},
	journal = J_IEEE_TSP,
	title = {Bi-directional Beamformer Training for Dynamic {TDD} Networks},
	year = {2018},
	volume = {66},
	number = {23},
	pages = {6252--6267},
    month= dec
    }

@article{Raz13,
author = {Razaviyayn, Meisam and Hong, Mingyi and Luo, Zhi-Quan},
title = {A Unified Convergence Analysis of Block Successive Minimization Methods for Nonsmooth Optimization},
journal = {SIAM Journal on Optimization},
volume = {23},
number = {2},
pages = {1126--1153},
year = {2013},
doi = {10.1137/120891009} }
